\documentclass[conference]{IEEEtran}
\IEEEoverridecommandlockouts

\usepackage[nocompress]{cite}
\usepackage{amsmath,amsfonts,amsthm}
\usepackage{algorithmic}
\usepackage{graphicx}
\usepackage{svg}
\usepackage{textcomp}
\usepackage{xcolor}
\usepackage{booktabs}
\usepackage{url}
\usepackage[justification=centering]{caption}
\usepackage{enumitem}
\usepackage{amsthm}
\usepackage{stfloats}
\usepackage{balance}
\usepackage{subfig}
\usepackage{threeparttable}
\usepackage[]{adjustbox}
\usepackage[ruled,noend]{algorithm2e}

\providecommand{\E}{\mathbb E}

\providecommand{\Description}[2][]{}

\newtheorem{theorem}{Theorem}[section]

\newtheorem*{remk*}{Remark}
\newtheorem{lemma}{Lemma}[section]

\newtheorem{proposition}{Proposition}[section]

\newcommand{\tabincell}[2]{\begin{tabular}{@{}#1@{}}#2\end{tabular}} 
\def\BibTeX{{\rm B\kern-.05em{\sc i\kern-.025em b}\kern-.08em
    T\kern-.1667em\lower.7ex\hbox{E}\kern-.125emX}}
\begin{document}

\title{Standardized Tolerance for Adaptive Reuse and Repair: A Method for HNSW Migration Under Embedding Model Change Scenarios
\thanks{This work was supported by the China Postdoctoral Science Foundation (Grant No. 2025M781464), and the Natural Science Foundation of Hebei Province of China (Grant No. F2026210032).}
}

\author{\IEEEauthorblockN{1\textsuperscript{st} Hengzhao Ma}
\IEEEauthorblockA{
\textit{Northeastern University}\\
Shenyang, China \\
mahz@neu.edu.cn}
\and
\IEEEauthorblockN{2\textsuperscript{nd} Chenyang Zhu}
\IEEEauthorblockA{
\textit{Northeastern University}\\
Shenyang, China \\
zhuchenyang@mails.neu.edu.cn}
\and
\IEEEauthorblockN{3\textsuperscript{rd} Bin Wang}
\IEEEauthorblockA{
\textit{Northeastern University}\\
Shenyang, China \\
binwang@mail.neu.edu.cn}
\and
\IEEEauthorblockN{4\textsuperscript{th} Xiaochun Yang}
\IEEEauthorblockA{
\textit{Northeastern University}\\
Shenyang, China \\
yangxc@mail.neu.edu.cn}

}

\maketitle

\begin{abstract}
When an embedding model changes, the same collection is re-encoded, and its vector index must be adapted to the new embedding space. Because the source and target vectors represent the same objects, suitable connections in the source index can be reused. We propose STARR (Standardized Tolerance for Adaptive Reuse and Repair) to migrate HNSW indexes after an embedding model change.
STARR compares standardized edge lengths over the same source neighbor lists in both embedding spaces and uses the correlation between their distance profiles to set the allowable increase in standardized edge length. 
It reconnects affected vertices by searching the partially migrated graph and shares the resulting candidates with their neighbors. 
We bound edge rejection under local distance distortion, characterize migration work, and prove that a final repair pass makes every vertex reachable from the entry point at the base layer while respecting degree limits. 
Single-threaded experiments on MS MARCO, Wikipedia, and CC3M, each with one million objects, show that STARR achieves 5.25--44.54$\times$ construction speedups over Fresh and migrates faster than both DGM variants on all 15 selected transitions. Recall@10 remains within one percentage point of Fresh in 10 cases, and query throughput exceeds DGM-Local in 14 cases. Parameter sweeps on three transitions further demonstrate favorable migration-time--recall tradeoffs beyond the default settings. Across all 240 directed transitions among 16 MS MARCO models, STARR achieves a median construction speedup of 16.71$\times$ with a median recall loss of 1.0 percentage point.
\end{abstract}


\section{Introduction}
\label{sec:introduction}
The growing demand for semantic retrieval has spurred extensive research into vector similarity search, which captures semantic similarity through distance-based neighborhood relationships in embedding spaces. Embedding models map multi-modal data, such as text, images, and audio, to dense, high-dimensional vectors, often yielding large vector collections. Efficient indexes are therefore essential for supporting vector similarity search at scale. In particular, graph-based vector indexes have attracted considerable attention for their strong empirical search performance~\cite{azizi2025graph,ootomo2024cagra,voruganti2025mirage}.

Recent research has devoted considerable effort to maintaining graph-based vector indexes in dynamic settings~\cite{gti2024,wolverine2025,prohnsw2026}, primarily focusing on the insertion, deletion, or update of individual vectors~\cite{greator2025,conda2026,odinann2026}. These operations, however, address only one aspect of index maintenance. 
Because vectors are derived from the underlying multi-modal objects through embedding models, an update to an individual vector must originate from an update to its underlying object. Another important source of change is the \emph{change of embedding models}, a common occurrence in real-world applications driven by version upgrades, changes in model size, or adaptation to new tasks~\cite{e5llm2024,mlembed2026,matryoshkaadaptor2024}. 

Such embedding model changes can transform the entire vector dataset even when the underlying  objects remain unchanged~\cite{continualcompatible2024,crossmodalcompatible2025,driftadapter2025}, requiring the index to adapt to a new similarity structure residing in an entirely new embedding space. This dataset-wide transformation presents a distinct maintenance challenge: techniques designed for individual vector updates do not directly address the problem of adapting the index to an entirely new vector dataset.

A straightforward solution is to rebuild the index from scratch on the new vectors, but this is prohibitively expensive, as graph construction is well known to be costly~\cite{wang2025flash,yang2025revisiting}. Rebuilding also overlooks a potential opportunity: although a change of embedding model produces two distinct sets of vectors, each vector in the old dataset has a counterpart in the new dataset derived from the same underlying object.
Because embeddings encode semantic relationships among these objects~\cite{platonic2024}, objects whose embeddings are close in the old space are likely to remain close in the new space. Therefore, an edge in the old index may still be useful for linking the corresponding vectors in the new index. 
This intuition motivates our central hypothesis: reusing suitable parts of the old graph-structured index can reduce the cost of constructing an index for the new vectors.

\begin{table}[t]
	\caption{Recall@10 (\%) on MS MARCO with one million objects and
		GTE-large as the target model.}
	\label{tab:direct-reuse-motivation}
	\centering
	\small
	\setlength{\tabcolsep}{4pt}
	\renewcommand{\arraystretch}{1.1}
	\begin{tabular*}{\columnwidth}{@{\extracolsep{\fill}}lrr@{}}
		\toprule
		Source model & \textbf{Fresh} & \textbf{Direct Reuse}\\
		\midrule
		BGE-large-en-v1.5 & 99.8 & 99.4\\
		Contriever & 99.8 & 96.8\\
		MPNet-base & 99.8 & 86.1\\
		MiniLM-L6-H384 & 99.8 & 75.5\\
		\bottomrule
	\end{tabular*}
\end{table}

To examine how well an old graph supports search after a model
change, we conduct a direct-reuse experiment on one million
MS MARCO objects. We fix GTE-large as the target model and use
old indexes built with four different embedding models. 
Direct Reuse replaces each node's vector with
its new counterpart while preserving identical edges. We compare its Recall@10 with
Fresh, an index built from scratch on the new vectors, using
the same query set and search parameters.
Table~\ref{tab:direct-reuse-motivation} shows that Fresh achieves
99.8\% recall, whereas Direct Reuse's recall ranges from  75.5\% to 99.4\%.
Thus, 
an old graph can retain some of its search capability on the new vectors,
but its suitability varies across model pairs, and copying all
connections can cause substantial recall loss. This motivates
identifying and retaining useful connections and repairing those
that no longer suit the target space, to achieve search quality
close to Fresh while avoiding a complete index rebuild.

\subsection{Problem and Challenges}
We now examine the task in greater detail. Let $X$ and $Y$ be vector datasets generated from the same set of underlying objects $O$ by different embedding models $M_X$ and $M_Y$, respectively. Given an existing graph-structured index $G_X$ over $X$, our goal is to construct a graph-structured index $G_Y$ over $Y$ by reusing suitable connections from $G_X$ and repairing those that no longer support effective search in the new embedding space. This task is called \emph{index migration} \cite{gu2026exploiting}.

Our preliminary experiments (Table \ref{tab:direct-reuse-motivation}) suggest that index migration involves two central decisions: which edges of the source index to retain, and how to add new edges so that the target index achieves search performance comparable to that of an index built from scratch. Addressing these decisions presents four key challenges. 

First, deciding whether to retain an edge $(i,j)$ from $G_X$ requires determining whether its endpoints remain near neighbors in the target embedding space. However, this neighborhood information must be obtained before an index over $Y$ is available, creating a circular dependency between neighborhood determination and index construction.

Second, the old and new vector datasets may differ in dimension and distance scale, so raw distances cannot be directly compared across the two spaces. Transferring neighborhood information therefore requires non-trivial techniques that account for these differences.

Third, any practical migration must be remarkably faster than rebuilding the index from scratch. 
For example, exhaustively testing all vector pairs to detect neighborhood changes would cost $O(n^2)$ time which defeats the purpose of migration and is prohibitive for large-scale datasets. 

Fourth, the migrated index must deliver query performance comparable to that of a freshly built index. Achieving construction savings while preserving query performance is essential to the practical value of index migration.

\begin{figure}[t]
	\centering
	\includegraphics[width=0.95\linewidth]{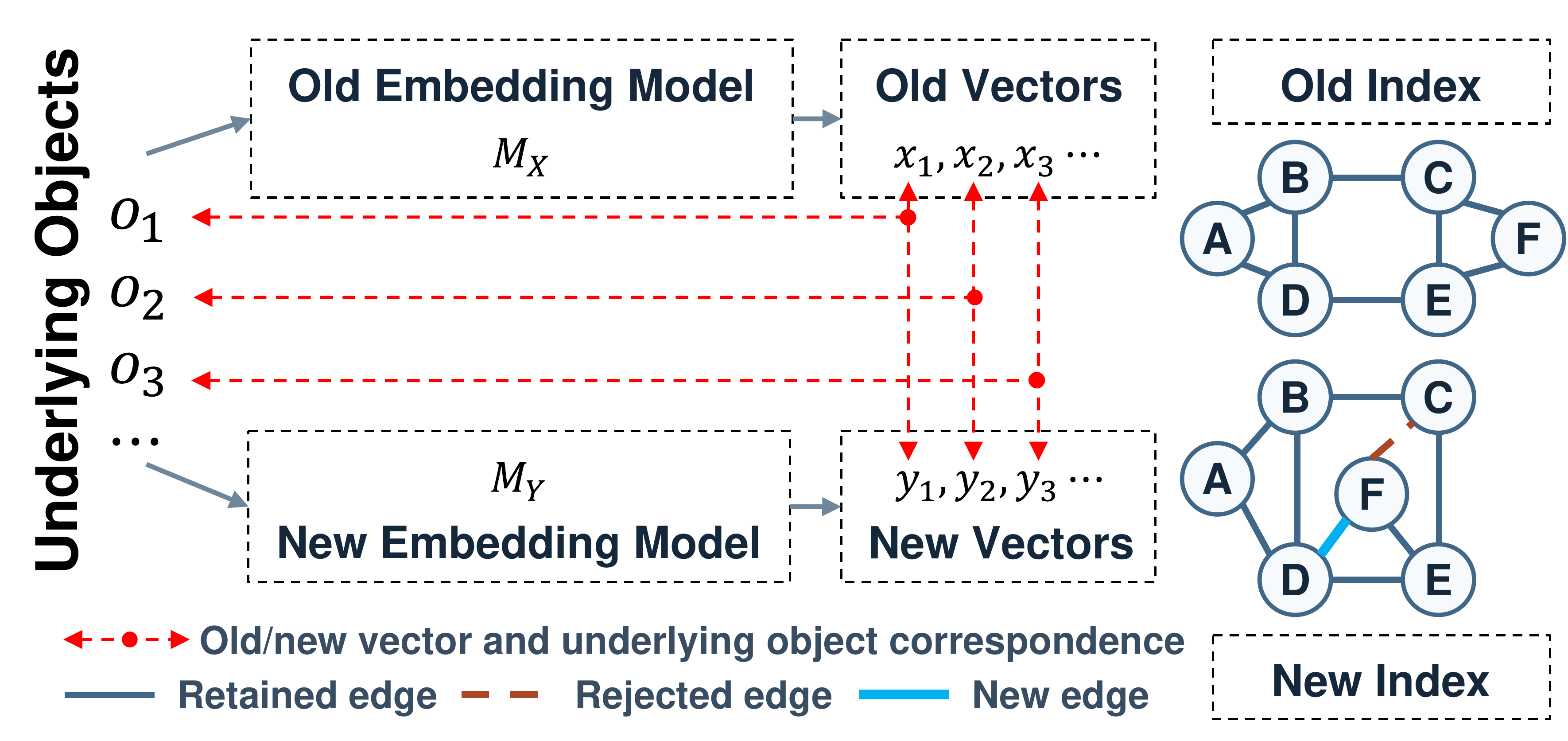}
	\caption{The idea of reuse-and-repair}
\end{figure}
\subsection{Approach and Contributions}
To address these challenges, we develop the following strategy based on the idea of reuse-and-repair.

First, we screen the old index $G_X$ using a novel criterion that distinguishes stable edges to retain from unstable edges to modify. This process requires no index over the new vector dataset $Y$, addressing the first challenge.

Second, the criterion is based on standardized distances, computed from the mean and variance of each node’s neighbor distance list. 
This normalization accounts for differences in dimension and distance scale across embedding spaces, addressing the second challenge.
Both criterion evaluation and edge screening operate within local neighbor lists, avoiding exhaustive pairwise comparisons and quadratic cost, thereby addressing the third challenge.

Third,  after a partial new index is obtained by screening and reusing the old index, we improve the new index through a reconnection stage, a local refinement pass, and a reachability repair stage. Together, these steps improve index quality and query performance, addressing the fourth challenge.

In summary, this paper makes the following contributions.
\begin{enumerate}[leftmargin=0em,itemindent=1.5em]
	\item We propose a lightweight neighborhood similarity metric that can be evaluated before the target index is built, allowing the identification of reusable edges without  quadratic cost.
	\item We develop an HNSW migration method that combines local edge
	screening, selective reconnection, candidate sharing, and reachability
	repair, which significantly reduces construction overhead compared to rebuilding from scratch.
	\item We analyze the time complexity of our method based on the careful inspection of the screening rules and repair process.
	\item We evaluate our method on various datasets against recent baseline methods. Across 15 transitions spanning different types of model change, STARR migrates faster than  baselines and achieves 5.25–44.54$\times$ construction speedups over Fresh, with Recall@10 within one percentage point of Fresh in 10 cases. 
    We further evaluate all 240 directed model transitions against Fresh to examine performance across embedding-model pairs. 
    Parameter sweeps and recall–throughput curves compare migration and query efficiency at comparable recall, while ablations quantify the effects of individual components on migration cost, recall, and reachability.
\end{enumerate}

\textcolor{black}{
	The rest of the paper is organized as follows. Section \ref{sec:preliminaries} introduces the preliminaries. Section \ref{sec:method} proposes the method for migration. Section \ref{sec:analysis} gives theoretical analysis of our method. Section \ref{sec:experiments} describes the experimental results. Section \ref{sec:rwork} reviews related work. Section \ref{sec:conclusion} concludes this paper.
}

\section{Preliminaries}
\label{sec:preliminaries}
\subsection{Objects, embeddings, and nearest neighbors}
\label{subsec:problem}
We place our research problem under the following setting.
Let $\mathcal O=\{o_i\}_{i=1}^{n}$ be a nonempty set of underlying objects,
with identifier set $\mathcal V=\{1,\ldots,n\}$. The old model $M_X$ maps $o_i$
to $x_i\in\mathbb R^{d_X}$, and the new model $M_Y$ maps it to
$y_i\in\mathbb R^{d_Y}$. We write the corresponding vector datasets as
$X=(x_i)_{i=1}^{n}$ and $Y=(y_i)_{i=1}^{n}$, and call them the source and target vector datasets, respectively. Their dimensions may differ, but
the identifier $i$ records the correspondence between representations.

For identifiers $i,j\in\mathcal V$, define
$D_X(i,j)=\|x_i-x_j\|_2, D_Y(i,j)=\|y_i-y_j\|_2$.
A query object $q$ has source representation $q_X$ and  target representation $q_Y$. Its distances to indexed objects are
$ D_X(q,i)=\|q_X-x_i\|_2, D_Y(q,i)=\|q_Y-y_i\|_2$.

For $A\in\{X,Y\}$, the $k$-nearest neighbors of $q_A$ in $A$, are the $k$ vectors in $A$ with minimum distances to $q_A$ under  $D_A$. Let $K_A(q)$ be the identifiers of the $k$ vectors, and let $r^A_{q,j}$ be the $j$-th
smallest distance. We use a fixed identifier order to break
ties on distance values.

Approximate $k$ nearest-neighbor (ANN) search instead finds a set of size $k$ that may not equal to $K_A(q)$, but with as much overlap as possible. The recall, which will be used in the experiments (Equation \ref{eq:recall}), is used to  measure the quality of the approximate nearest neighbor set.

\subsection{HNSW index}
\label{subsec:hnsw}
ANN search is usually supported by index structures. This paper focuses on HNSW index.

An HNSW index is a hierarchy of directed graphs $\{G_A^h\}=(\mathcal V_h,E_A^h)$, $h=0,\ldots,H$, where $H$ is the maximum level, and $A\in\{X,Y\}$ denotes the embedding space. 
Each vertex $i$ has a nonnegative integer height $h_i\le H$ and belongs to
layers $0,\ldots,h_i$. 
The level-0 layer, or the base layer, has $\mathcal V_0=\mathcal V$, and $\mathcal V_h=\{i:h_i\ge h\}$. 
The entry point $e\in\mathcal V_H$ is the designated starting vertex for hierarchical search.
The adjacency list $N_A^h(i)$ contains the outgoing neighbors of $i$
at layer $h$. 
The outgoing degree is denoted as $|N_A^h(i)|$.

HNSW constructs the index incrementally, assigning each new vertex
$i$ a random height $h_i$. Starting from $e$, it descends greedily
through layers above $h_i$, repeatedly moving to a closer neighbor
and using the closest vertex found as the next layer's entry point.
From layer $\min\{H,h_i\}$ to layer 0, it searches for candidates
with width \texttt{efConstruction}, selects up to $M$ neighbors
using a diversity heuristic, and adds outgoing and reverse
connections. The heuristic can reject a candidate that is closer
to an already selected neighbor than to the vertex being connected,
avoiding redundant connections. Lists exceeding their capacities
are pruned using this heuristic.
Higher layers contain exponentially fewer vertices, forming a
sparse navigation structure, while the base layer contains all
vertices for finer local exploration. The outgoing-list capacities
used here are $C_0=2M$ and $C_h=M$ for $h>0$, with common upper
bound $C=2M$; these limits do not bound incoming degree.
Queries exploit this hierarchy by descending greedily from the
top-layer entry point to approach the query's neighborhood.
At the base layer, best-first search expands the closest pending
vertex and maintains a bounded set of promising candidates,
from which the closest $k$ are returned as approximate nearest
neighbors. Query width \texttt{efSearch} controls the retained
candidate set,
trading query cost for search accuracy~\cite{hnsw}.

\subsection{The migration task}
\label{subsec:migration-task}
We now define the migration task to tackle in this paper.\\
\textbf{Input:}  source and target vector sets $X$ and $Y$, together with the source HNSW index $G_X$.\\
\textbf{Output:} an HNSW index $G_Y$ for ANN search on $Y$.\\
\textbf{Requirements:} Migration should be substantially faster than rebuilding the index from scratch and should yield query performance comparable to that of a freshly built index.

\section{Method Description}
\label{sec:method}
\SetKwFunction{MigMain}{MigrateHNSW}
\SetKwFunction{MigScreen}{ScreenEdges}
\SetKwFunction{MigReverse}{BatchReverse}
\SetKwFunction{MigReconnect}{ReconnectAndShare}
\SetKwFunction{MigSearch}{MultiSeedSearch}
\SetKwFunction{MigReach}{EnsureReachable}
\SetKwFunction{MigAttach}{TryAttach}
\SetKwFunction{MigFlood}{BFS}
\SetKwFunction{MigGreedy}{GreedySearch}
\SetKwFunction{MigConnect}{ConnectHNSW}
\SetKwFunction{MigPrune}{DiversityPrune}
\SetKwFunction{MigBridgeSearch}{BaseSearch}
\SetKwFunction{MigVisit}{Visit}
\SetKwProg{MigProc}{Procedure}{}{end}
\DontPrintSemicolon
\LinesNumbered
\SetAlgoNoEnd
\SetAlgoLined
\SetAlFnt{\small}
\SetAlCapFnt{\small}
\SetAlCapNameFnt{\small}
\SetAlgoNlRelativeSize{-1}

Given source and target vectors for the same objects,  and a source HNSW index, our method STARR, standing for  \underline{S}tandardized \underline{T}olerance for \underline{A}daptive \underline{R}euse and \underline{R}epair,
constructs the target index in three stages. 
The overall algorithm is given in Algorithm~\ref{alg:mig:main}.

First, we screen source edges by comparing local distance relationships in the two embedding spaces using the same source neighbor lists. 
We retain edges for which these relationships change only slightly and mark vertices for repair if they lose any outgoing edges or have empty neighbor lists.

Next, we search the current target graph for candidate neighbors and use them to reconnect vertices in the repair set. We also share these candidates with nearby vertices, allowing vertices outside the repair set to gain connections. When the repair set is small, selected reverse
connections are added to retained edges before reconnection. 

Finally, we check base-layer reachability from the entry point
and add bridges to unreached portions of the graph to ensure the full graph reachability.

By reusing
source connections and the search routes of the source index, the method
concentrates reconnection searches on affected vertices to reduce
construction cost while improving search quality of  the migrated target index.

\begingroup
\SetKwFunction{MigReconnect}{Reconnect}
\LinesNumbered
\SetAlgoNoEnd
\SetAlgoLined
\SetAlFnt{\small}
\SetAlCapFnt{\small}
\SetAlCapNameFnt{\small}
\SetAlgoNlRelativeSize{-1}

\begin{algorithm}[t]
	\caption{HNSW index migration}
	\label{alg:mig:main}
	\SetInd{0.3em}{0.6em}
	\KwIn{Source index $G_X$ over $X$; target vectors $Y$;
		tolerance $\tau$; search width $B$}
	\KwOut{Target index $G_Y$}
	Initialize $G_Y$ with $Y$, source metadata, and empty lists\;
	$\mathcal U\gets\emptyset$\;
	\ForEach{$i\in\mathcal V$ and $h=0,\ldots,h_i$}{
		$N_Y^h(i)\gets S_i^h$ using Eq.~\eqref{eq:stable-set}\;
		\If{$S_i^h\ne N_X^h(i)$ or $(S_i^h=\emptyset\ \text{and}\ n>1)$}{
			$\mathcal U\gets\mathcal U\cup\{i\}$\;
		}
	}
	\If{$|\mathcal U|(C_0+1)<n$}{
		Batch stable reverse connections (Section~\ref{subsec:reverse})\;
	}
	\ForEach{$i\in\mathcal U$ in decreasing height, then increasing ID}{
		$\mathcal C_i\gets$ \MigReconnect{$G_Y,i,B$}
		(Section~\ref{subsec:reconnect})\;
		\ForEach{$v\in N_Y^0(i)$ with $|N_Y^0(v)|<C_0$}{
			$A\gets\mathcal C_i\setminus(\{v\}\cup N_Y^0(v))$\;
			\If{$A\ne\emptyset$}{
				Append $\mathop{\mathrm{arg\,min}}_{c\in A}D_Y(v,c)$ to $N_Y^0(v)$\;
			}
		}
	}
	Repair entry-point reachability (Section~\ref{subsec:reachability})\;
	\Return $G_Y$\;
\end{algorithm}
\endgroup

\subsection{Initializing the Target Index}
\label{subsec:method:initialization}

To initialize the target index, we  allocate  the vertex
storage with target vector dimension $d_Y$, preserving each vertex's height $h_i$, 
and write the corresponding target vectors $y_i$. 
The vertex count, entry point $e$, maximum level $H$, and neighbor-list capacities retain
their source settings, while every neighbor list is initialized
as empty.
The initialized target index therefore contains all target vectors
and preserves the source index's object correspondence and layer
membership, but has no edges. The next stage examines source edges
and adds only those satisfying the reuse criterion to the target
index.

\subsection{Detecting Reusable Edges}
\label{subsec:edge-stability}
Throughout migration, $N_X^h(i)$
denotes the fixed source neighbor list, while $N_Y^h(i)$ denotes
the current target neighbor list.
We assess each source edge by the length change relative to
the other edges in the same neighbor list. Since the source and
target spaces may differ in dimension and distance scale, we
compare standardized distances rather than raw edge lengths.
For each nonempty source list $N_X^h(i)$  and $A\in\{X,Y\}$, the
mean and variance of the distances are
\begin{align}
	\mu^A_{i,h}
	&=\frac{1}{|N_X^h(i)|}\sum_{j\in N_X^h(i)}D_A(i,j), \label{eq:mean}\\
	(\sigma^A_{i,h})^2
	&=\frac{1}{|N_X^h(i)|}\sum_{j\in N_X^h(i)}
	(D_A(i,j)-\mu^A_{i,h})^2.
	\label{eq:variance}
\end{align}
We call $\{D_A(i,j)\mid j\in N_X^h(i)\}$ the \emph{distance
	profile} in space $A$. Each edge length is standardized as
\begin{equation}
	z^A_{i,h}(j)=
	\begin{cases}
		\displaystyle\frac{D_A(i,j)-\mu^A_{i,h}}{\sigma^A_{i,h}},
		&\sigma^A_{i,h}>0,\\
		0,   &\sigma^A_{i,h}=0.
	\end{cases}
	\label{eq:zscore}
\end{equation}
The resulting z-score describes the edge's relative length within
its local distance profile. The zero-variance convention covers
singleton and constant-distance lists.

To adapt screening to the preservation of the local distance
profile, we measure agreement between the source and target
z-score lists:
\begin{equation}
	\rho_{i,h}=\frac{1}{|N_X^h(i)|}
	\sum_{j\in N_X^h(i)}z^X_{i,h}(j)z^Y_{i,h}(j).
	\label{eq:correlation}
\end{equation}
When both variances are positive, $\rho_{i,h}$ is the Pearson
correlation between the profiles. If either variance is zero,
the convention above gives $\rho_{i,h}=0$.

Given a tolerance parameter $\tau\ge0$, we define the allowed
increase in standardized edge length as
\begin{equation}
	t_{i,h}=\tau\times\min\{1,\max\{0,\rho_{i,h}\}\}.
	\label{eq:tolerance}
\end{equation}
This scales the threshold between zero and $\tau$: strong positive
agreement permits a larger increase, while weaker agreement makes
screening more restrictive. The change of standardized distance for an individual edge is
\begin{equation}
	\Delta_{i,h}(j)=z^Y_{i,h}(j)-z^X_{i,h}(j).
	\label{eq:zchange}
\end{equation}
We retain neighbors whose standardized edge lengths do not
increase beyond this threshold:
\begin{equation}
	S_i^h=\{j\in N_X^h(i):\Delta_{i,h}(j)\le t_{i,h}\}.
	\label{eq:stable-set}
\end{equation}
The criterion is one-sided: an edge whose standardized length
decreases is retained, while an increase is accepted only within
the correlation-adjusted tolerance. It therefore distinguishes
relative improvement from deterioration rather than treating both
as distortion. All comparisons use existing source neighbor lists,
so screening requires neither a target-space index nor an
exhaustive comparison of all vector pairs. The resulting sets
$S_i^h$ identify the source directions selected for initial reuse.

\subsection{Adding Stable Reverse Connections}
\label{subsec:reverse}
Using the sets identified in Section~\ref{subsec:edge-stability},
we add each retained edge $i\to j$, $j\in S_i^h$, to the target
index at layer $h$, without automatically adding its reverse. 
Vertices that lose an outgoing edge at any layer, or have an empty
neighbor list at an occupied layer when $n>1$, form the
\emph{repair set} $\mathcal U$, with $u=|\mathcal U|$.
It remains fixed throughout subsequent repair.

When the repair set is small, reconnection and candidate sharing
may update only a limited portion of the base graph. We therefore
first supplement retained edges with reverse connections if
$u(C_0+1)<n$.
Here, $C_0+1$ bounds the number of base-layer lists that reconnection
and sharing can update per repaired vertex: its own list and at
most $C_0$ neighboring lists.

The reverse pass operates layer by layer. A retained edge $i\to j$
proposes $j\to i$ only if that direction is missing and was not
explicitly rejected during screening. After collecting all proposals
for a layer, we merge each recipient's proposals with its current
neighbors. We write the union directly if it fits within $C_h$;
otherwise, we apply diversity pruning once before writing it.
This batch update avoids repeated pruning and does not guarantee
bidirectional connections. The resulting graph provides the search
routes for reconnecting vertices in $\mathcal U$.

\begin{algorithm}[t]
	\caption{Adding stable reverse connections}
	\label{alg:mig:reverse}
	\SetInd{0.3em}{0.6em}
	\KwIn{Screened target graph $G_Y$; rejected directions $R^h$}
	\KwOut{Updated target adjacency lists; unchanged repair set}
	\MigProc{\MigReverse{$G_Y,\{R^h\}_{h=0}^{H}$}}{
		\For{$h\gets0$ \KwTo $H$}{
			$P_v\gets\emptyset$ for every $v\in\mathcal V_h$\;
			\ForEach{$i\in\mathcal V_h$ in increasing internal-ID order}{
				\ForEach{$j\in N_Y^h(i)$}{
					\If{$i\notin N_Y^h(j)$ and $(j,i)\notin R^h$}{
						Append $i$ to $P_j$\;
					}
				}
			}
			\ForEach{$v\in\mathcal V_h$ with $P_v\ne\emptyset$}{
				$L\gets N_Y^h(v)$ followed by $P_v$\;
				\If{$|L|>C_h$}{
					$L\gets$ \MigPrune{$v,L,C_h$}\;
				}
				$N_Y^h(v)\gets L$\;
			}
		}
	}
\end{algorithm}
\subsection{Search-Based Connection Repair}
\label{subsec:reconnect}
The preceding stages partially establish target index connections, but vertices in $\mathcal U$ may still lack suitable neighbors.
 We therefore use each vertex's target
vector as a query to search the current target graph and repair its
connections. 

For each $i\in\mathcal U$, greedy search starts at $e$ and descends
through layers above $h_i$ to find an entry close to $y_i$.
We process layers $h_i,\ldots,0$ in descending order,
so repaired upper-layer connections can guide subsequent searches.
Before searching each layer, we use $i$'s neighbor list
$N_Y^h(i)$ in the current target graph and an
entry vertex to seed best-first search. At layer $h_i$, the
entry comes from the upper-layer greedy traversal; at each lower
layer, it comes from neighbor selection at the preceding layer.
The repair search width $B$ limits the retained set to at most
$B$ candidate neighbors, excluding $i$. All searches use the graph updated by earlier repairs.

We apply the HNSW diversity heuristic \cite{hnsw} to select neighbors from the
candidate set, update $i$'s outgoing connections, prune lists exceeding their
capacities, and attempt reverse
insertion into the selected neighbors' lists. 
If the updated neighbor list for $i$ has free slots, we append missing neighbors from the saved
list until all are restored or capacity is reached. Thus, restoration
uses neighbors already present in the target graph immediately
before this layer's search, including connections retained or added
by preceding operations. 

This reconnection process updates an existing vertex's connections while preserving
its identifier and height. An edge rejected during screening may
be rediscovered and inserted, as long as it is eligible for  reconnection. The base-layer search candidates
also support the candidate sharing described next.

\subsection{Sharing Search Candidates}
\label{subsec:sharing}
The preceding reconnection stage initiates searches only for vertices
in $\mathcal U$. However, a vertex that retains all its source edges
may still lack useful neighbors in the target space. We therefore
share repair search candidates with nearby vertices, allowing them
to gain connections without launching independent searches, improving index quality at low cost.

For each repaired vertex $i$, we save the candidate set $\mathcal C_i$
retained by its base-layer search before diversity pruning, with
$|\mathcal C_i|\le B$. This set includes candidates that may
subsequently be pruned from $i$'s list. Local neighborhoods can
overlap, making these candidates worth testing as neighbors of
$i$'s neighbors. For example, a candidate $c$ close to another
neighbor $v$ may be considered redundant when selecting neighbors
for $i$, yet still provide a useful direct connection from $v$.

After reconnecting $i$ at the base layer and restoring its pre-search
neighbors, we examine its current neighbors
$v\in N_Y^0(i)$. For each $v$ with spare capacity, we exclude $v$
itself and its existing neighbors from the shared candidates:
\begin{equation}
	A_{i,v}=\mathcal C_i\setminus(\{v\}\cup N_Y^0(v)).
	\label{eq:eligible-sharing}
\end{equation}
If $A_{i,v}$ is nonempty, we select the candidate $c^*_{i,v}$ closest to $v$
in the target space,
\begin{equation}
	c^*_{i,v}=\mathop{\mathrm{arg\,min}}_{c\in A_{i,v}}D_Y(v,c)
	\label{eq:sharing}
\end{equation}
and add the directed edge $v\to c^*_{i,v}$.

Each recipient gains at most one edge per repaired vertex $i$.
We skip recipients whose lists are full or whose eligible candidate
sets are empty. Sharing preserves existing edges and adds no reverse
connection. By reusing search results, it extends connection updates
beyond $\mathcal U$ without requiring a separate graph search for each recipient.

\subsection{Ensuring Reachability from the Entry Point}
\label{subsec:reachability}
Filtering and pruning can leave some vertices unreachable from the
entry point $e$. After reconnection and candidate sharing, we therefore
check base-layer reachability and add bridges where needed.

We first perform a breadth-first traversal from $e$, marking reached
vertices and recording the edge that first discovers each vertex.
These edges form a traversal tree and are protected during subsequent
updates to preserve existing paths from the entry point.
For each unreached vertex $r$, a base-layer best-first search from $e$
with query $y_r$ finds reached vertices as candidate bridge sources,
or \emph{anchors}. We try them in increasing distance to $r$.
An anchor with a free slot accepts an edge to $r$ directly; a full
anchor instead replaces its farthest outgoing non-tree edge.
If none of the search candidates can accept a bridge, we scan the
remaining reached vertices for a suitable anchor.

After adding a bridge, we include it in the traversal tree and resume
traversal from $r$ to mark newly reached vertices. We repeat this
process until every vertex is reachable from $e$. All updates respect
the base-layer neighbor capacity, while tree-edge protection preserves
previously established paths.


\begin{algorithm}[t]
	\caption{Ensuring Reachability from the Entry Point}
	\SetInd{0.3em}{0.6em}
	\label{alg:mig:reachability}
	\KwIn{Current target graph $G_Y$; entry point $e$; width $B$}
	\KwOut{Every vertex reachable from $e$ at layer 0, within capacity}
	\MigProc{\MigReach{$G_Y,B$}}{
		Run base-layer BFS from $e$; let $R$ be the reached vertices
		and $T$ their protected discovery-tree edges\;
		\ForEach{$r\in\mathcal V$ in increasing internal-ID order}{
			\If{$r\notin R$}{
				Search layer 0 from $e$ with query $y_r$ and width $B$\;
				Choose the closest returned anchor $a$ with a free slot
				or an outgoing edge outside $T$\;
				\If{no such anchor was returned}{
					Scan $R$ in BFS discovery order for an anchor satisfying
					the same condition\;
				}
				\If{$|N_Y^0(a)|=C_0$}{
					Remove $a$'s farthest outgoing edge outside $T$\;
				}
				Add $a\to r$ to $G_Y$ and $T$, and add $r$ to $R$\;
				Resume BFS from $r$, extending $R$ and protecting each
				new discovery edge in $T$\;
			}
		}
	}
\end{algorithm}

\section{Analysis}
\label{sec:analysis}
Throughout the analysis we use the following notation. There are $n\ge1$ paired objects with Euclidean distances
$D_X(i,j)$ and $D_Y(i,j)$ in dimensions $d_X$ and $d_Y$.
The source list $N_X^h(i)$ contains outgoing neighbors at layer $h$;
vertex $i$ occupies layers $0,\ldots,h_i$. Layer $0$ contains every
vertex. The entry point is $e$, and outgoing capacities satisfy
$1\le C_h\le C$. The repair set $\mathcal U$ is fixed after screening,
with $u=|\mathcal U|$.

\subsection{Local Screening and Repair Population}
\label{ana:subsec:screening}
In this part we focus on the screening process on a local neighbor list of length $s$.
Based on the definition of distance profiles, standardized distance change $\Delta_j$ and tolerance $t$, let 
\begin{equation}
	\begin{split}
		\alpha=s^{-1}|\{j:\Delta_j>t\}|
	\end{split}
	\label{ana:eq:screen-rule}
\end{equation}
which is the rejected  fraction of the screening process.

First, if the distances are positively affinely transformed, i.e., there exists $\gamma>0$ and $\delta$ such that $D_Y(i,j)=\gamma D_X(i,j) +\delta$ for each $j$, the standardized distance profiles are identical and all neighbors survive. We then give a bound on $\alpha$ as follows.

\begin{lemma}[Score identities and rejected-edge bound]
	\label{ana:lem:score}
	Let $\chi_A=\mathbf1\{\sigma_A>0\}$. For every nonempty list,
	\begin{equation}
		s^{-1}\sum_j\Delta_j=0,\qquad
		v:=s^{-1}\sum_j\Delta_j^2=\chi_X+\chi_Y-2\rho.
		\label{ana:eq:identity}
	\end{equation}
	If $v=0$, no edge is rejected. If $t>0$,
	\begin{equation}
		\alpha\le\frac{v}{v+t^2}
		=\frac{2(1-\rho)}{2(1-\rho)+\tau^2\rho^2}.
		\label{ana:eq:cantelli}
	\end{equation}
	 If $t=0<v$, the bound $\alpha\le1$ remains valid.
\end{lemma}
\begin{proof}
	Standardization gives, including the zero-variance case,
	\begin{equation}
		s^{-1}\sum_jz_j^A=0,\qquad
		s^{-1}\sum_j(z_j^A)^2=\chi_A.
		\label{ana:eq:prof-standardized}
	\end{equation}
	Hence $s^{-1}\sum_j\Delta_j=0$, and expanding the square yields
	\begin{equation}
		v=s^{-1}\sum_j(z_j^Y-z_j^X)^2=\chi_Y+\chi_X-2\rho.
	\end{equation}
	If $v=0$, every $\Delta_j=0\le t$, so all edges survive.
	For $t>0$, let $Z$ be a uniformly chosen $\Delta_j$; then
	$\E Z=0$, $\E Z^2=v$, and $\alpha=\Pr(Z>t)$. For any $b\ge0$,
	\begin{equation}
		\alpha\le\frac{\E[(Z+b)^2]}{(t+b)^2}
		=\frac{v+b^2}{(t+b)^2}.
		\label{ana:eq:prof-cantelli-raw}
	\end{equation}
	Taking $b=v/t$ gives $\alpha\le v/(v+t^2)$. Since $t>0$ implies
	$\rho>0$ and $\chi_X=\chi_Y=1$, substituting
	$v=2(1-\rho)$ and $t=\tau\rho$ proves \eqref{ana:eq:cantelli}.
	The case $t=0<v$ follows from $\alpha\le1$.
\end{proof}

As a side note, singleton lists always survive, and no nonempty list can become
empty under exact screening, since its score changes sum to zero.

\begin{theorem}[Listwise screening stability]
	\label{ana:thm:filter}
	Consider a source neighbor list of node $i$ with $\sigma_X>0$. Suppose for every entry of this list there are constants 	$0<\ell\le L$ satisfying
	\begin{equation}
		\ell D_X(i,j)\le D_Y(i,j)\le L D_X(i,j)
		\label{eq:bilipschitz}
	\end{equation}

	Set $\varepsilon=\frac{L-\ell}{L+\ell}$, $	r=\varepsilon\sqrt{1+(\mu_X/\sigma_X)^2}$.
	If $r<1$, then $\sigma_Y>0$ and
	$\rho\ge\rho_*:=\sqrt{1-r^2}$.
	For $\tau>0$, the rejected fraction satisfies
	\begin{equation}
		\alpha\le
		\frac{2(1-\rho_*)}{2(1-\rho_*)+\tau^2\rho_*^2}.
		\label{ana:eq:distfilter}
	\end{equation}
	For $\tau\ge0$, a sufficient condition for every edge to survive is
	\begin{equation}
		\sqrt{2s(1-\rho_*)}\le\tau\rho_*.
		\label{ana:eq:allkeep}
	\end{equation}
\end{theorem}
\begin{proof}
	Let $a$ and $c$ be the old and new distance vectors of the source list.
	Let $\gamma=(L+\ell)/2$ and write $c/\gamma=a+e$; then
	\eqref{eq:bilipschitz} gives $|e_j|\le\varepsilon a_j$.
	Let $P$ subtract the list mean and use
	$\|w\|_{\rm rms}^2=s^{-1}\sum_jw_j^2$. For $x=Pa$, $f=Pe$,
	and $y=Pc/\gamma=x+f$,
	\[
	\|f\|_{\rm rms}\le\varepsilon\|a\|_{\rm rms}
	=\varepsilon\sqrt{\mu_X^2+\sigma_X^2}=r\|x\|_{\rm rms}.
	\]
	Since $r<1$, $y\ne0$, so $\sigma_Y>0$ and $\rho$ is the cosine
	between $x$ and $y$. Decompose
	$f/\|x\|_{\rm rms}=\beta x/\|x\|_{\rm rms}+g$, where $g\perp x$.
	Then $\beta^2+\|g\|_{\rm rms}^2\le r^2$ and
	\begin{equation}
		\begin{split}
			\rho&=\frac{1+\beta}{\sqrt{(1+\beta)^2+\|g\|_{\rm rms}^2}}\\
			&\ge\frac{1+\beta}{\sqrt{1+2\beta+r^2}}
			\ge\sqrt{1-r^2}=\rho_*.
		\end{split}
		\label{ana:eq:prof-rho-of-t}
	\end{equation}
	The last inequality follows from $(\beta+r^2)^2\ge0$ and
	$1+\beta>0$. For $\tau>0$, the bound in \eqref{ana:eq:cantelli}
	decreases for $0<\rho\le1$, giving \eqref{ana:eq:distfilter}.
	Finally, \eqref{ana:eq:identity} gives
	$|\Delta_j|\le\sqrt{sv}
	\le\sqrt{2s(1-\rho_*)}$.
	Under \eqref{ana:eq:allkeep}, this is at most
	$\tau\rho_*\le\tau\rho=t$, so every edge survives.
\end{proof}

\subsection{From rejected edges to repaired vertices.}
\label{subsec:rejected-vertexes}
\begin{lemma}
	Let $R$ count rejected directed edges across all layers. For $n>1$,
	let $z_0$ count vertices with an originally empty list at an occupied
	layer; set $z_0=0$ for $n=1$. By the definition of the repair set $\mathcal U$, we can bound  $u = |\mathcal U| $ by 
	\begin{equation}
		u\le\min\{n,z_0+R\},\qquad
		R\le\sum_{i,h}s_{i,h}\alpha^{\rm ub}_{i,h}.
		\label{ana:eq:repairpopulation}
	\end{equation}
	Set $\alpha^{\rm ub}=0$ when $v=0$; otherwise use
	\eqref{ana:eq:distfilter} only if $\tau>0$ and $r<1$, and use $1$
	in all remaining cases. 	
\end{lemma}
\begin{proof}
Every vertex in $\mathcal U$ either contributes to $z_0$ or is
the tail of a rejected edge, so $u\le\min\{n,z_0+R\}$.
The second inequality follows by summing
$\alpha_{i,h}\le\alpha^{\rm ub}_{i,h}$, weighted by $s_{i,h}$,
over the nonempty lists.
\end{proof}

\subsection{Capacity-Preserving Reachability}
\label{ana:subsec:reach}
Section~\ref{subsec:rejected-vertexes} bounded how many vertices end up in
the repair set $\mathcal U$. The next result is a 
structural guarantee about that procedure's output: whatever size
$\mathcal U$ turns out to have, repair leaves the base layer connected
from the entry point, without ever asking any vertex to hold more
outgoing edges than its capacity allows.

\begin{proposition}[Rooted reachability]
	\label{ana:prop:reachability}
	Let $\mathcal R_0$ be the base-layer vertices reachable from $e$
	before reachability repair. Protecting traversal-tree
	edges and attaching unreached vertices through free or unprotected
	slots terminates with every vertex reachable from $e$, respects
	outgoing capacities, and adds at most $n-|\mathcal R_0|$ bridges.
\end{proposition}
\begin{proof}
	Maintain a protected spanning out-tree $T$ of the reached set
	$\mathcal R$, initially $\mathcal R_0$. Since protected edges are
	never removed, all vertices in $\mathcal R$ remain reachable from $e$.
	If $\mathcal R\ne\mathcal V$, any leaf of $T$ has no protected
	outgoing edge. Its capacity is at least $1$, so it has either a free
	slot or an unprotected edge that can be replaced by a bridge to an
	unreached vertex. Thus the fallback scan always finds an eligible
	anchor if the search does not. Adding and protecting the bridge,
	then extending $T$ by traversal from its endpoint, preserves
	capacities and existing protected paths while increasing
	$|\mathcal R|$ by at least one. At most
	$n-|\mathcal R_0|$ bridges suffice to reach every vertex.
\end{proof}

\subsection{Total Time Complexity}
\label{ana:subsec:time}
\newcommand{\lgp}[1]{\ell\!\left(#1\right)}
\setlength{\emergencystretch}{2em}

	\begin{theorem}\label{thm:time-final}
		Let $M\ge1$ be the HNSW neighbor parameter, 
        yielding neighbor list capacity $2M$ at the base layer and $M$ at higher layers.  
        Let the maximum height of HNSW be $H=\max_i h_i$. 
        Let $\Lambda=\sum_i(h_i+1)$, $m=\sum_{i,h}|N_X^h(i)|\le C\Lambda$.
        Let $\mathcal U$ be the repair set, $u=|\mathcal U|$, and
		$A=\sum_{i\in\mathcal U}(h_i+1)$. 
        Let $B\ge M$ be the  search width.
        Assume that $M$ and $B$ are fixed,
		$\Lambda=O(n)$, and $A=O(H+1+u\lgp{n})$. The last condition holds whenever
		$H=O(\lgp{n})$. Denote $\lgp{x}=\log_2(2+x)$. 
		
		Let $r_0$ be the number of vertices reachable from the entry point in the
		base layer immediately before
		reachability repair. If reachability repair inserts $b$ bridges, take any
		integer $\bar b$ with $b\le\bar b\le n-r_0$, and set
		\[
		J_{\bar b}=\frac{\bar b(2n-\bar b-1)}{2}.
		\]
	 For these counts, the worst-case migration
		time satisfies
		\[
		T_{\rm mig}=
		O\bigl(n(d_X+d_Y)+n[H+1+\lgp{n}]
		+(un+J_{\bar b})(d_Y+\lgp{n})\bigr)
		\]
	\end{theorem}
	
	\begin{proof}
		\emph{Initialization and screening.}
		Validating and copying the target vectors costs $O(nd_Y)$, and
		initializing the inherited neighbor lists costs $O(\Lambda)$ for fixed
		$M$. The number $m$ of directed source edges is at most $2M\Lambda=O(n)$.
		Screening evaluates the source and target distances on those edges and
		computes list statistics and retention decisions in linear time in the
		number of entries. It flags vertices that lose an outgoing edge or have
		an empty occupied source list when $n>1$.  These operations cost $O(\Lambda+m(d_X+d_Y))$ in total, so the
		initialization and screening cost is $O(n(d_X+d_Y))$.
		
		\emph{Batch reversal and scheduling.}
		The batch reversal scans all $n$ identifiers at every layer and
		makes at most one reverse proposal per retained edge. The merged pools
		contain at most $2m$ entries in total because the retained lists and the
		reverse proposals each contribute at most $m$. An overflowing pool of
		size $s$ takes $O(s[d_Y+\lgp{n}])$ time to score and prune. Each candidate
		is compared against at most $2M$ selected neighbors, and its heap
		operations cost $O(\lgp{n})$ because $s=O(n)$. Membership and rejection
		checks take constant time per retained edge. Building the
		rejection hash sets costs $O(H+1+m)$. The reversal cost is
		therefore $O(n(H+1)+n[d_Y+\lgp{n}])$. Forming the repair list and
		scanning its $u$ entries at each height costs $O(n+u(H+1))$. Together
		these stages add $O(n[H+1+\lgp{n}])$ beyond the $O(nd_Y)$ already
		allowed in the first term.
		
		\emph{Connection repair and sharing.}
		Let $n_h$ be the number of vertices occupying layer $h$, so
		$\sum_{h=0}^H n_h=\Lambda$. Each flagged vertex uses greedy search above
		its height and one beam search at each occupied layer. Greedy search  scans at most
		$n_h$ distinct lists at layer $h$. A beam call marks each discovered
		vertex once and expands it at most once. The fixed outgoing capacities
		give $O(n_h)$ distance evaluations and adjacency checks, including the
		seeds. The search makes $O(n_h)$ heap operations, each costing
		$O(\lgp{n})$ because at most $n_h\le n$ entries can be pending. With $B$
		and $M$ fixed, selecting neighbors, updating their reverse lists, restoring
		saved entries, and
		sharing base-layer candidates take $O(d_Y+1)$ per repaired vertex-layer
		pair. Thus all repairs take $O(u\Lambda[d_Y+\lgp{n}])=
		O(un[d_Y+\lgp{n}])$.
		
		\emph{Reachability repair.}

	The initial breadth-first traversal and all resumed traversals take $O(n)$ total time.
	By Proposition~\ref{ana:prop:reachability}, each bridge increases the
	reached set and $b\le n-r_0$.	
		Let $r_j$ be the reached-set size before bridge $j$. The reached set is
		closed under outgoing edges at that point, so a beam search from the
		entry point visits at most $r_j$ vertices. Sorting the at most $B$
		returned anchors takes constant time. The fallback scans at most $r_j$
		anchors, each with a list of bounded size. Search and attachment
		therefore cost $O(r_j[d_Y+\lgp{n}])$ per bridge. The $r_j$ are distinct
		positive integers less than $n$, and the largest $\bar b$ such integers
		sum to $J_{\bar b}$. Hence the bridge work is
		$O(J_{\bar b}[d_Y+\lgp{n}])$.
	\end{proof}

\section{Experiments}
\label{sec:experiments}
\subsection{Experimental Setup}
\label{subsec:experiment-protocol}
\subsubsection{Datasets and model transitions}
We evaluate HNSW migration on three datasets, each containing one
million objects. MS MARCO~\cite{msmarco} and Wikipedia use text
embeddings, while CC3M uses image embeddings.
For each transition, the source and target embeddings represent the
same objects. Our main comparison includes five transitions per
dataset, spanning fine-tuning, version updates, model scaling, and
changes in model family or architecture.
Table~\ref{tab:experiment-transitions} lists these 15 transitions and
their embedding dimensions. To examine performance across a broader
range of model pairs, we also evaluate all $A_{16}^{2}$ directed
transitions among 16 models on MS MARCO.

\subsubsection{Compared methods}
We compare STARR (code available \cite{STARR-Github}) with Fresh, DGM-Local, and
DGM-Search~\cite{gu2026exploiting}. Fresh builds a new HNSW index on the
target embeddings. While both DGM variants reuse
the source graph to construct the target index,  
DGM-Local obtains candidates through local graph
expansion, and DGM-Search uses hop-bounded beam search. We adapt the public
DGM implementation to the HNSW hierarchy, using the same degree
limits and L2 distance metric as the other methods.

\subsubsection{Implementation and parameters}
All experiments run on an Intel Core Ultra 7 265K server, compiled with \texttt{-O2}
and FP32 vectors using single-threaded C++17 implementations. Source and Fresh indexes use $M=16$ and
\texttt{efConstruction}$=200$, with degree limits of 32 at the base
layer and 16 at upper layers. Our default parameters are $\tau=2$
and $B=16$. DGM-Local uses mean-centered position sign (MPS)
screening with a candidate budget of 400. DGM-Search uses a beam
capacity of 256 and a four-hop limit. Both DGM variants use
$\alpha=1.1$. We set \texttt{efSearch}$=200$ for the main comparison
and vary it on fixed indexes to compare recall and query throughput.
We jointly vary $\tau$ and $B$ in the sensitivity experiments.

\begin{table}[t]
	\caption{Selected model transitions. BGE labels omit \texttt{-en}.
		The suffix \texttt{ft} denotes MS MARCO fine-tuning.}
	\label{tab:experiment-transitions}
	\centering
	\footnotesize
	\setlength{\tabcolsep}{3pt}
	\begin{tabular}{@{}llr@{\(\;\to\;\)}ll@{}}
		\toprule
		Source model & Target model & \multicolumn{2}{c}{$d_X\to d_Y$} & Scenario\\
		\midrule
		\multicolumn{5}{@{}l@{}}{\textbf{MS MARCO and Wikipedia}}\\
		BGE-base-v1.5   & BGE-base-v1.5-ft & 768 & 768  & Fine-tuning\\
		BGE-base        & BGE-base-v1.5    & 768 & 768  & Version update\\
		BGE-small-v1.5  & BGE-large-v1.5   & 384 & 1024 & Scale up\\
		MiniLM-L6-H384  & GTE-large        & 384 & 1024 & Cross-family\\
		\addlinespace[2pt]
		\multicolumn{5}{@{}l@{}}{\textbf{MS MARCO only}}\\
		BGE-large-v1.5  & GTE-large        & 1024 & 1024 & Cross-family\\
		\addlinespace[2pt]
		\multicolumn{5}{@{}l@{}}{\textbf{Wikipedia only}}\\
		E5-large-v2     & BGE-base-v1.5    & 1024 & 768 & Cross-family\\
		\addlinespace[2pt]
		\multicolumn{5}{@{}l@{}}{\textbf{CC3M}}\\
		DINOv1-ViT-S/16 & DINOv1-ViT-B/16  & 384  & 768 & Scale up\\
		DINOv1-ViT-B/16 & DINOv2-base      & 768  & 768 & Generation change\\
		CLIP-ViT-L/14   & SigLIP-base      & 768  & 768 & Cross-family\\
		SigLIP-base     & CLIP-ViT-L/14    & 768  & 768 & Cross-family\\
		DINOv1-ResNet50 & DINOv2-base      & 2048 & 768 & Cross-architecture\\
		\bottomrule
	\end{tabular}
\end{table}

\subsubsection{Evaluation protocol and metrics}
All methods use the same 1,000 indexed objects as queries for each
dataset, with self-matches included. We compute exact top-10
neighbors in the target embedding space as ground truth.
For query set $\mathcal Q$, let $T_k(q)$ and $A_k(q)$ denote the
exact and returned neighbor sets, respectively. Recall is
\begin{equation}
	\mathrm{Recall@}k=
	\frac{1}{|\mathcal Q|}
	\sum_{q\in\mathcal Q}
	\frac{|A_k(q)\cap T_k(q)|}{k},
	\label{eq:recall}
\end{equation}
where $k=10$. We measure query throughput in queries per second
(QPS) after warm-up. Recall--QPS curves report median QPS over five
timed runs per configuration.
We report target-index construction or migration time and define
construction speedup as $T_{\mathrm{Fresh}}/T_{\mathrm{method}}$.
Migration time includes target-index initialization and graph
migration. Source-index construction, embedding generation, disk
I/O, and ground-truth computation are excluded. Construction and
migration times are measured once per configuration.

\subsection{Overall Performance}
\label{subsec:overall-performance}
\begin{figure*}[t]
	\centering
	\includegraphics[width=\textwidth]{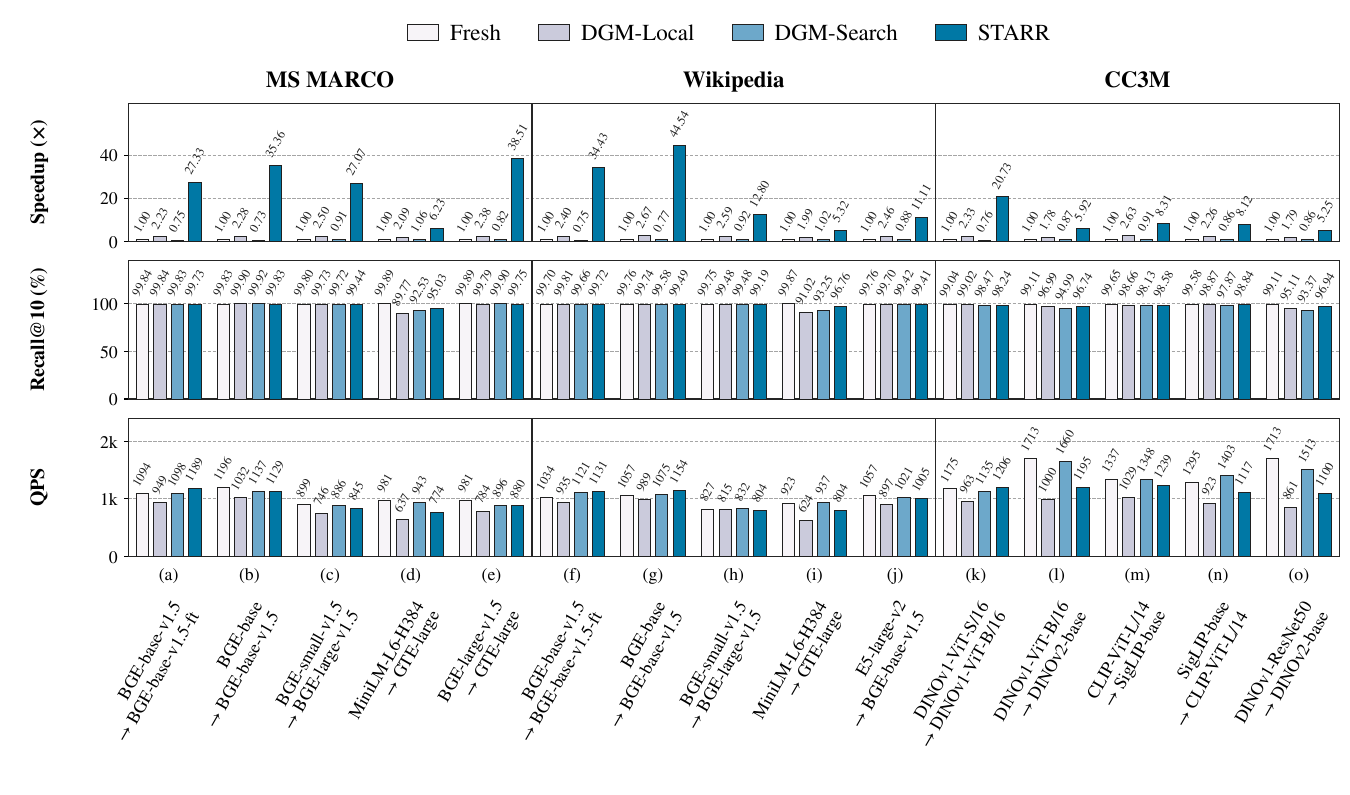}
	\caption{Performance on 15 model transitions.
		Dataset blocks are arranged by column, and rows show construction
		speedup over Fresh, Recall@10, and QPS at \texttt{efSearch}$=200$.
		Higher values are better for all three metrics.
		\texttt{ft} denotes fine-tuning.}
	\label{fig:overall-performance}
	\Description{Grouped bar charts compare four methods across five
		transitions each on MS MARCO, Wikipedia, and CC3M. STARR has the
		highest construction speedup in all 15 cases. Its recall is within
		one percentage point of Fresh in 10 cases and exceeds both DGM
		variants on the two MiniLM-to-GTE transitions. Its QPS exceeds
		DGM-Local in 14 cases.}
\end{figure*}

Figure~\ref{fig:overall-performance} compares construction speedup,
Recall@10, and QPS under single-threaded execution across 15 model
transitions on the three datasets. The cases span both text and image embeddings,
including updates within a model family and cross-family transitions. STARR has the lowest
migration time in every case, achieving $5.25$--$44.54\times$
speedups over Fresh. DGM-Local achieves $1.78$--$2.67\times$
speedups, while DGM-Search takes longer than Fresh in 13 cases.

On most transitions, STARR maintains recall close to Fresh
and the two DGM variants. Recall@10 is within one percentage point
of Fresh in 10 cases. For BGE-base to BGE-base-v1.5 on MS MARCO,
both STARR and Fresh achieve 99.83\% recall, with STARR achieving a $35.36\times$
construction speedup. On Wikipedia, the same transition yields
the largest speedup, $44.54\times$, with a recall loss of only
0.27 percentage points. These model-update cases show that
migration can substantially reduce construction cost while
preserving search quality close to that of a fresh index.

STARR also achieves higher recall than both DGM variants
on several difficult transitions. For MiniLM to GTE-large,
Recall@10 reaches 95.03\% on MS MARCO and 96.76\% on Wikipedia.
These values exceed DGM-Local by 5.26 and 5.74 percentage points,
and DGM-Search by 2.50 and 3.51 percentage points, respectively,
while migration remains faster than either variant. On CC3M,
DINOv1-ResNet50 to DINOv2-base likewise achieves higher recall
than both DGM variants. A gap to Fresh remains in these cases,
highlighting the remaining quality loss under the default settings.

For query throughput, STARR exceeds DGM-Local in 14 cases
and matches or exceeds Fresh on some transitions. DGM-Search
provides higher QPS than STARR in most cases, but also incurs
substantially higher migration costs. STARR combines lower
migration cost with competitive query throughput, although some
difficult transitions incur a throughput loss relative to Fresh.
For DINOv1-ResNet50 to DINOv2-base, for example, QPS decreases
from approximately 1,713 to 1,100, alongside a $5.25\times$
construction speedup.

\subsection{Performance across parameter configurations}
\begin{figure*}[t]
 \centering
 \includegraphics[width=\textwidth]{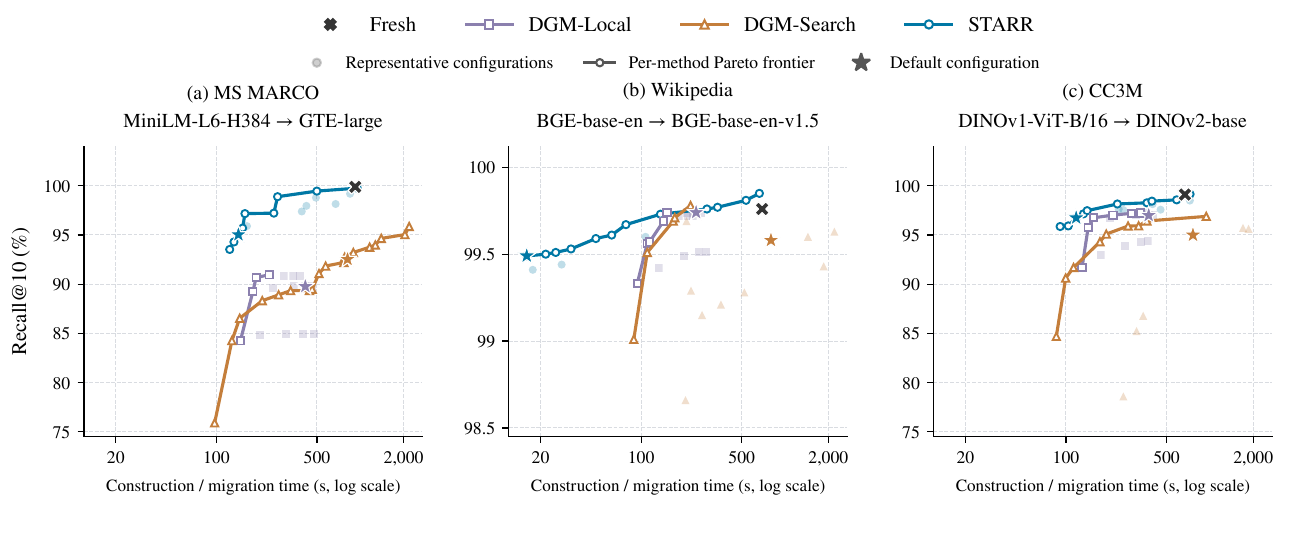}
 \caption{Construction--recall tradeoffs under parameter sweeps on
 three model transitions, evaluated with 1,000 queries.
 Faint markers show representative configurations, and lines connect
 the nondominated configurations from each method's full measured
 grid. Stars mark default configurations, and dark crosses
 show Fresh. The time axis is logarithmic.}
 \label{fig:construction-recall-tradeoff}
 \Description{Three panels compare STARR, DGM-Local, DGM-Search,
 and Fresh on MS MARCO, Wikipedia, and CC3M. STARR offers lower
 migration costs at shared recall thresholds and reaches higher
 recall than both DGM variants on the two difficult transitions.
 Its default configurations lie on the measured Pareto frontiers
 at the low-cost end.}
\end{figure*}

Figure~\ref{fig:construction-recall-tradeoff} compares the
construction--recall tradeoffs of the three migration methods on
three representative transitions, using 1,000 queries and fixed
\texttt{efSearch}$=200$. We evaluate 28 STARR, 15 DGM-Local, and
48 DGM-Search configurations per transition. The curves connect
each method's measured Pareto frontier, minimizing migration time
and maximizing Recall@10. Representative configurations are shown
to reduce clutter, with all frontier points and defaults retained.

STARR offers a favorable time--recall tradeoff across the scanned
configurations. On MS MARCO, the fastest configurations reaching
at least 95\% recall take 142.35 seconds for STARR and 2,043.21
seconds for DGM-Search, a $14.35\times$ difference. DGM-Local
does not reach this threshold within the scanned grid. On
Wikipedia, reaching at least 99.5\% recall takes 21.72 seconds
with STARR, compared with 110.11 and 110.33 seconds for DGM-Local
and DGM-Search. On CC3M, STARR reaches 98.15\% recall in 227.66
seconds, while neither DGM variant reaches 98\% in the scan.
STARR's default configuration lies on the measured Pareto frontier
at the low-cost end in all three transitions. Its advantage thus
extends beyond the default comparison, while additional repair
provides options for higher recall requirements.

\subsection{Query Performance}
\label{subsec:query-performance}
\begin{figure*}[t]
 \centering
 \includegraphics[width=\textwidth]{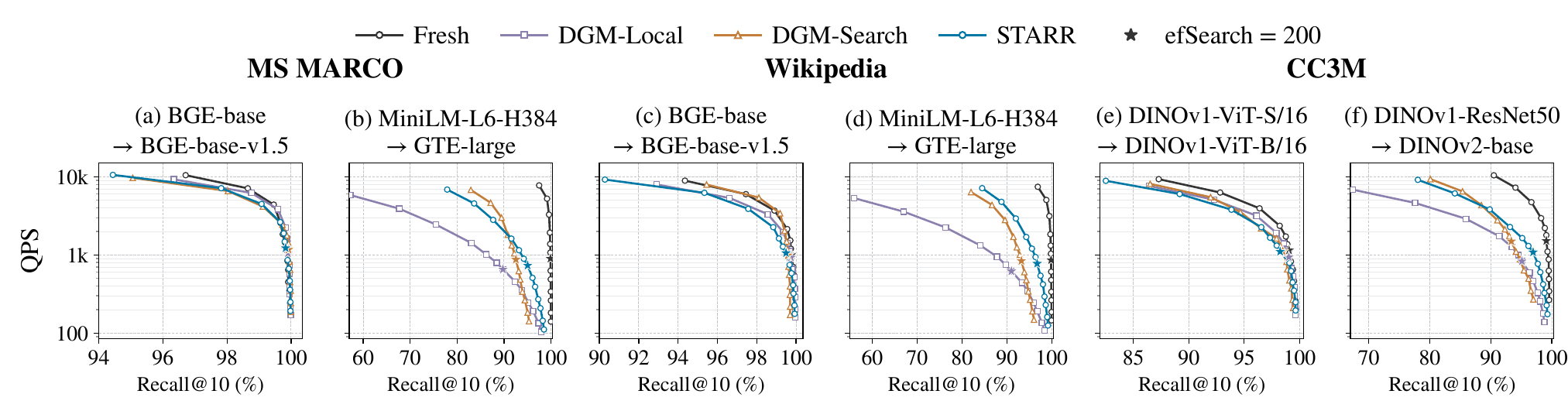}
 \caption{Recall--QPS curves on six model transitions. Each curve connects 13 measured configurations in increasing \texttt{efSearch} order, from 10 to 1600, on a fixed index. Reported QPS is the median of five timed runs, plotted on a log scale. Stars mark \texttt{efSearch}$=200$. Panels (a,b), (c,d), and (e,f) correspond to MS MARCO, Wikipedia, and CC3M, respectively.}
 \label{fig:recall-qps}
 \Description{Six panels compare Fresh, DGM-Local, DGM-Search,
 and STARR in one row. Panels a and b show a BGE version update
 and MiniLM to GTE-large on MS MARCO. Panels c and d show the
 same transitions on Wikipedia. Panels e and f show DINOv1
 model scaling and DINOv1-ResNet50 to DINOv2-base on CC3M.
 Recall is on the horizontal axis and QPS on the vertical axis.}
\end{figure*}

Figure~\ref{fig:recall-qps} compares query throughput across six
transitions as \texttt{efSearch} varies on fixed indexes. For each
recall threshold, we compare the highest measured QPS among the
configurations that meet or exceed it.

For BGE version updates, relative query efficiency varies by
dataset. At the 99\% recall threshold on MS MARCO, STARR
achieves approximately 4,487 QPS, outperforming all three baselines
in query throughput. On Wikipedia, DGM-Search
provides higher throughput for the same transition.

STARR's advantage over the DGM variants is more pronounced on the
difficult MiniLM-to-GTE-large transitions. At the 95\% recall
threshold, STARR achieves approximately 733 QPS on MS MARCO
and 1,203 QPS on Wikipedia. These rates are $2.97\times$ and
$4.95\times$ those of DGM-Local, and $3.99\times$ and $4.50\times$
those of DGM-Search, respectively. Thus, our migrated indexes
require less query time to meet the same recall requirement
on these transitions.

A similar advantage appears for DINOv1-ResNet50 to DINOv2-base
on CC3M. At the 98\% recall threshold, STARR achieves
approximately 601 QPS, $2.36\times$ that of DGM-Local.
DGM-Search reaches a maximum recall of 96.99\% within the tested
range. Fresh still offers higher throughput on these difficult
transitions, revealing a remaining query-performance gap on hard cross-architecture
migrations.

\subsection{Performance Across Model Transitions}
\label{subsec:model-transitions}
\begin{figure*}[t]
 \centering
 \includegraphics[width=\textwidth]{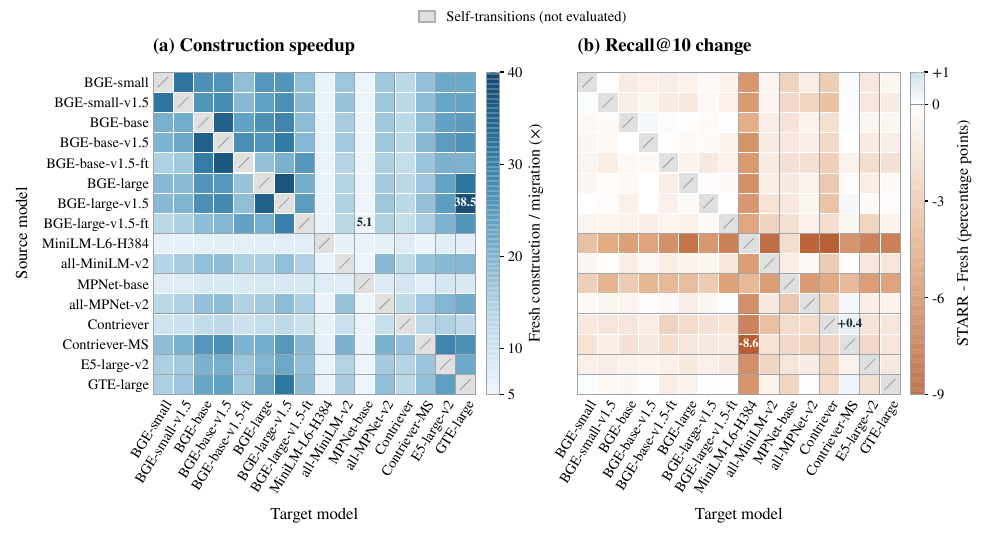}
 \caption{Migration performance across all 240 directed model
 transitions on MS MARCO with one million objects.  Both panels use
 the same source-model rows, labeled at left, and target-model
 columns.  Gray diagonal cells are
 unevaluated self-transitions. (a) Fresh construction time
 divided by migration time. (b) Recall@10 of STARR minus that of
 Fresh, in percentage points; orange indicates a decrease, white
 no change, and pale blue an increase. Selected extrema are annotated.
 }
 \label{fig:msmarco-matrix}
 \Description{Two aligned 16-by-16 heatmaps of all directed
 MS MARCO model transitions except self-transitions. Source-model
 labels are shared at left, and each panel has a vertical color bar.
 Construction
 speedups range from 5.10 to 38.51 times. Recall changes range
 from minus 8.6 to plus 0.4 percentage points, with larger losses
 concentrated in several source rows and target columns.}
\end{figure*}

We evaluate all $A_{16}^{2}$ directed model transitions on
MS MARCO. Figure~\ref{fig:msmarco-matrix} shows construction
speedup and Recall@10 changes relative to Fresh. In both panels the 
rows are source models and columns are target models.

Migration is faster than Fresh in all 240 transitions, with
construction speedups of $5.10$--$38.51\times$ and a median of
$16.71\times$. These savings do not always preserve recall close
to that of Fresh. Recall@10 changes range from $-8.6$ to $+0.4$
percentage points, with a median of $-1.0$ percentage point.
In 121 transitions, recall is within one percentage point of
Fresh or higher; 19 transitions match or exceed Fresh.
Thus, construction savings extend across the full matrix,
whereas the resulting search quality depends on the model pair.

The heatmaps also show that both construction speedup and recall loss
vary substantially across source models, even when the migration
parameters are held fixed. For transitions starting from MiniLM-L6-H384
and MPNet-base, the median recall losses relative to Fresh are 5.9 and
3.9 percentage points, respectively. The corresponding median construction
speedups are $6.60\times$ and $8.29\times$. For transitions starting from
BGE-large-v1.5, the median recall loss is only 0.3 percentage points,
while the median construction speedup reaches $19.20\times$.

Reversing the migration direction can change performance outcome. Migrating
from Contriever to Contriever-MS yields a $12.89\times$ speedup,
with Recall@10 of 99.5\% compared with 99.1\% for Fresh.
The reverse transition achieves a $17.41\times$ speedup, but
recall decreases from 98.3\% for Fresh to 96.1\% after migration.
Model transitions must therefore be evaluated directionally:
performance in one direction does not determine performance
in the reverse direction.

All transitions in this subsection use the same tolerance,
$\tau=2$, which need not suit every model pair. When source
and target models differ substantially in their semantic spaces
and neighborhood structure, a permissive tolerance may retain
too many edges that no longer suit the target space, reducing
recall. For such transitions, we recommend a smaller $\tau$:
stricter edge screening triggers more reconnection work,
trading additional migration cost for higher recall.
We examine this tradeoff further in the parameter sensitivity
experiments.

\subsection{Ablation Study}
\label{subsec:ablation}
Table~\ref{tab:ablation-main} compares STARR with six ablated
variants on six transitions, using one million
objects, 1,000 queries, $\tau=2$, and $B=16$. We disable correlation
scaling, restoration of pre-search neighbors, candidate sharing,
reachability repair, or batch reverse insertion individually,
and also disable sharing and reachability repair together,
giving 42 configurations.

\begin{table*}[t]
	\caption{Ablations on one million objects per dataset with
		$\tau=2$, $B=16$, and 1,000 queries. Time is migration time in
		seconds, and Recall is Recall@10 in percent. Each ablation disables the indicated component or components of STARR.
		Reachability refers to the final entry-point reachability repair.}
	\label{tab:ablation-main}
	\centering
	\footnotesize
	\setlength{\tabcolsep}{2.3pt}
	\renewcommand{\arraystretch}{1.18}
	\begin{tabular*}{\textwidth}{@{\extracolsep{\fill}}l*{12}{r}@{}}
		\toprule
		Variant & \multicolumn{4}{c}{MS MARCO} &
		\multicolumn{4}{c}{Wikipedia} & \multicolumn{4}{c}{CC3M}\\
		\cmidrule(lr){2-5}\cmidrule(lr){6-9}\cmidrule(lr){10-13}
		& \multicolumn{2}{c}{\tabincell{c}{ BGE-base-en $\rightarrow$\\
		BGE-base-en-v1.5}} &
		\multicolumn{2}{c}{\tabincell{c}{MiniLM-L6-H384\\
		$\rightarrow$ GTE-large}} &
		\multicolumn{2}{c}{\tabincell{c}{BGE-base-en $\rightarrow$\\
		BGE-base-en-v1.5}} &
		\multicolumn{2}{c}{\tabincell{c}{MiniLM-L6-H384\\
		$\rightarrow$ GTE-large}} &
		\multicolumn{2}{c}{\tabincell{c}{DINOv1-ViT-B/16 \\$\rightarrow$ DINOv2-base}} &
		\multicolumn{2}{c}{\tabincell{c}{CLIP-ViT-L/14\\ $\rightarrow$ SigLIP-base}}\\
		\cmidrule(lr){2-3}\cmidrule(lr){4-5}\cmidrule(lr){6-7}
		\cmidrule(lr){8-9}\cmidrule(lr){10-11}\cmidrule(lr){12-13}
		& Time $\downarrow$ & Recall $\uparrow$
		& Time $\downarrow$ & Recall $\uparrow$
		& Time $\downarrow$ & Recall $\uparrow$
		& Time $\downarrow$ & Recall $\uparrow$
		& Time $\downarrow$ & Recall $\uparrow$
		& Time $\downarrow$ & Recall $\uparrow$\\
		\midrule
		\textbf{STARR}
		&23.23&99.83&154.64&95.03&17.56&99.49&154.73&96.76&129.51&96.74&108.42&98.58\\
		w/o correlation scaling
		&21.93&99.81&99.19&89.22&17.79&99.39&108.87&90.88&70.02&94.47&55.48&97.17\\
		w/o neighbor restoration
		&23.05&99.81&95.89&90.39&16.13&99.39&93.17&92.90&82.52&91.82&73.29&96.11\\
		w/o candidate sharing
		&23.05&99.81&120.86&93.74&16.12&99.46&119.62&95.33&103.95&96.28&86.65&98.31\\
		w/o reachability repair
		&23.01&99.83&154.15&95.03&17.29&99.49&158.32&96.75&129.17&96.72&107.75&98.54\\
		w/o sharing + reachability
		&22.79&99.81&120.45&93.73&15.57&99.46&118.48&95.32&103.76&96.26&86.20&98.28\\
		w/o batch reverse insertion
		&12.44&99.84&154.28&95.03&17.23&99.49&157.95&96.76&129.23&96.74&108.27&98.58\\
		\bottomrule
	\end{tabular*}
\end{table*}

Correlation scaling and neighbor restoration provide the largest
recall gains on the difficult transitions. Removing correlation
scaling lowers Recall@10 by 5.81 and 5.88 percentage points for
MiniLM to GTE on MS MARCO and Wikipedia, respectively. On
MS MARCO, migration time falls from 154.64 to 99.19 seconds,
reflecting the cost of the additional repair induced by adaptive
screening. Disabling neighbor restoration reduces recall by
4.64 percentage points on the same transition and by 4.92 points
for DINOv1 to DINOv2 on CC3M. Both components have smaller
recall effects on BGE version updates.

Candidate sharing improves recall across all six transitions,
with gains ranging from 0.02 to 1.43 percentage points.
For MiniLM to GTE on MS MARCO, disabling sharing reduces
migration time from 154.64 to 120.86 seconds and recall from
95.03\% to 93.74\%. Sharing thus trades additional migration
work for better search quality.

\begin{table}[t]
	\caption{Vertices unreachable from the entry point after migration. 
		STARR includes final reachability repair. The w/o repair variant
		disables this step.
		}
	\label{tab:ablation-reachability}
	\centering
	\small
	\setlength{\tabcolsep}{1.8pt}
	\begin{tabular*}{\columnwidth}{@{\extracolsep{\fill}}llcr@{}}
		\toprule
		Dataset & Transition & STARR & \tabincell{c}{w/o \\repair}\\
		\midrule
		MS MARCO & BGE-base-en $\rightarrow$ BGE-base-en-v1.5 & 0 & 2\\
		& MiniLM-L6-H384 $\rightarrow$ GTE-large & 0 & 305\\
		Wikipedia & BGE-base-en $\rightarrow$ BGE-base-en-v1.5 & 0 & 85\\
		& MiniLM-L6-H384 $\rightarrow$ GTE-large & 0 & 222\\
		CC3M & DINOv1-ViT-B/16 $\rightarrow$ DINOv2-base & 0 & 854\\
		& CLIP-ViT-L/14 $\rightarrow$ SigLIP-base & 0 & 1,825\\
		\bottomrule
	\end{tabular*}
\end{table}

Reachability repair primarily improves structural coverage.
Table~\ref{tab:ablation-reachability} shows that STARR reaches
all one million vertices in every case, while disabling repair
leaves 2--1,825 vertices unreachable. The corresponding recall
loss is only 0--0.04 percentage points. Disabling sharing and
repair together produces recall close to disabling sharing alone,
but also leaves unreachable vertices. Recall therefore captures
only part of the benefit of entry-point reachability repair.

Batch reverse insertion is triggered only for the MS MARCO BGE
update among the six complete STARR configurations. Disabling it
reduces migration time from 23.23 to 12.44 seconds, with essentially
unchanged recall, 99.83\% versus 99.84\%. Recall is identical in
the other five transitions, where the gate remains inactive.
The Wikipedia BGE update also activates this pass when correlation
scaling is disabled, showing that an ablation can change the
execution of another component through the gate condition.

\subsection{Parameter Sensitivity}
\label{subsec:parameter-sensitivity}
Tolerance $\tau$ and repair search width $B$ control STARR's
repair effort. A smaller $\tau$ tightens edge screening, while a
larger $B$ expands the repair search. The sweeps in
Figure~\ref{fig:construction-recall-tradeoff} combine
$\tau\in\{0,0.5,1,1.5,2,2.5,3\}$ with
$B\in\{16,32,64,128\}$. For the DGM comparison, we scan
DGM-Local's candidate budget $C\in\{64,128,256,400,800\}$
and DGM-Search's beam capacity $L\in\{64,128,256,512\}$
and hop limit $H\in\{2,3,4,6\}$. Both DGM methods use
$\alpha\in\{1.0,1.1,1.2\}$.

For MiniLM to GTE-large on MS MARCO, stricter screening and wider repair search substantially improve recall. Moving from the default $(\tau,B)=(2,16)$ to $(0,32)$ raises Recall@10 from 95.03\% to 98.89\%, with migration time increasing from 142.35 to 266.35 seconds. At $(0,64)$, STARR reaches 99.46\% recall in 499.18 seconds, versus Fresh at 99.89\% in 922.23 seconds. Further increasing $B$ to 128 at $\tau=0$ yields 99.76\% recall in 952.26 seconds, slightly exceeding Fresh’s construction time. This illustrates the increasing cost of approaching Fresh recall. On CC3M, increasing $B$ from 16 to 64 at $\tau=2$ raises recall from 96.74\% to 98.15\%, with migration time increasing from 117.76 to 324.71 seconds.

Sensitivity near the default depends on the transition. On
Wikipedia, recall varies only from 99.43\% to 99.51\% for $\tau\in\{1.5,2,2.5\}$ at $B=16$. Over the same tolerance range, recall varies from
94.29\% to 95.70\% on MS MARCO and from 95.92\% to 97.15\%
on CC3M. Thus, the default configuration provides an efficient starting point across
the three transitions, while stricter screening or wider repair
search is useful when a difficult transition requires higher recall.


\section{Related Work}
\label{sec:rwork}

\subsection{Graph-based ANN indexing.}
HNSW constructs a hierarchy of proximity graphs through incremental
insertion, graph search, and diversity-based neighbor selection~\cite{hnsw}.
Subsequent work reduces graph construction cost through hardware and
algorithmic improvements. CAGRA develops highly parallel graph
construction and search for GPUs~\cite{ootomo2024cagra}, while Flash
uses compact vector codes to accelerate distance comparisons on
modern CPUs~\cite{wang2025flash}. Yang et~al. revisit proximity-graph
construction and introduce a construction framework with a new pruning
strategy~\cite{yang2025revisiting}. These approaches improve the
construction process over supplied vectors. Our work examines an
additional source of savings when an index already exists for another
representation of the same objects: transferring suitable graph
connections and concentrating searches on affected vertices.

\subsection{Dynamic vector indexes.}
Dynamic ANN methods maintain search structures as vectors arrive or
are removed. GTI combines a tree with a lightweight graph to support
efficient insertions and deletions~\cite{gti2024}. Wolverine repairs
monotonic search paths disrupted by deletions~\cite{wolverine2025},
and Greator localizes small-batch updates to reduce graph-maintenance
I/O and redundant neighbor pruning~\cite{greator2025}. CONDA uses
connectivity-aware pruning, bidirectional link reinforcement, and lazy
deletion to preserve search quality under continuous
updates~\cite{conda2026}. OdinANN directly inserts vectors into an
on-disk graph to stabilize search performance during
ingestion~\cite{odinann2026}. These methods share our interest in
reusing an existing graph and repairing its connections. The migration
setting introduces a different reuse decision: the object identifiers
remain fixed, but a model change replaces the entire collection's
representations and can change their dimensionality. Processing these
changes as individual deletions and insertions does not itself identify
which source connections remain useful under target-space distances.
Our screening rule makes this decision using paired local distance
profiles before a target index is available.

\subsection{Embedding compatibility and model upgrades.}
Compatible representation learning addresses model changes at the representation level. 
Continual variants align new and historical gallery features~\cite{continualcompatible2024}, cross-modal variants project embeddings into legacy spaces~\cite{crossmodalcompatible2025}, Drift-Adapter maps new queries into the old space~\cite{driftadapter2025}, and Embedding-Converter converts embeddings between models~\cite{embeddingconverter2025}. Our setting assumes that target vectors are already available and
addresses index construction under target-space distances.
Representation conversion
and index migration thus address different stages of a model upgrade.

\subsection{Cross-model graph index migration}
Gu et al. \cite{gu2026exploiting} studied the same index migration task and proposed the two DGM variants evaluated here. DGM-Local expands inherited neighborhoods to generate candidates for each vertex, then evaluates and prunes them in the target space. DGM-Search uses hop-bounded beam traversal to discover candidates for each vertex. By contrast, STARR compares standardized edge lengths across the two embedding spaces and uses the correlation of the paired profiles to set the retention threshold. A vertex enters the repair set if screening rejects any of its outgoing edges or if it has an empty list at an occupied layer. This confines reconnection searches to the flagged vertices, while candidate sharing can also update nearby vertices outside the repair set.

\section{Conclusion}\label{sec:conclusion}
Useful neighborhood structure can survive an embedding-model change and reduce the cost of constructing the target index. STARR identifies reusable edges through standardized distance profiles and repairs affected neighborhoods to obtain a reachable HNSW graph. Experiments show substantial construction savings across model transitions, with recall close to that of Fresh in many cases. The parameter sweeps show that migration quality depends on the model pair and repair budget. Stricter screening and wider repair search can improve recall on difficult transitions at additional cost. Future work will focus on improving the navigability of migrated graphs and extending reuse and repair to other graph-based indexes.
\bibliographystyle{IEEEtranS}
\bibliography{library}

@misc{STARR-Github,
  author    = {Chenyang Zhu and Hengzhao Ma},
  title     = {{STARR GitHub} Repository},
  url       = {https://github.com/QuaintAunt/STARR},
  year      = {2026}
}

@article{gu2026exploiting,
  title={Exploiting Residual Reachability for Cross-Model Migration of Graph-Based Indexes in Approximate Nearest Neighbor Search},
  author={Gu, Baoyuan and Zhong, Xiaoyao and Jin, Jiabao and Cheng, Peng and Ni, Wangze and Li, Haotian and Song, Jingkuan and Shen, Heng Tao},
  journal={arXiv preprint arXiv:2609.23036},
  year={2026}
}

@misc{hnsw,
  author    = {{Yu.} A. Malkov and D. A. Yashunin},
  title     = {Efficient and robust approximate nearest neighbor search using {Hierarchical Navigable Small World} graphs},
  note      = {arXiv:1603.09320, version 4},
  year      = {2018},
  url       = {https://arxiv.org/abs/1603.09320}
}

@misc{msmarco,
  author    = {Payal Bajaj and others},
  title     = {{MS MARCO}: A Human Generated {MAchine} Reading {COmprehension} Dataset},
  note      = {arXiv:1611.09268},
  year      = {2016},
  url       = {https://arxiv.org/abs/1611.09268}
}

@inproceedings{e5llm2024,
  author    = {Liang Wang and Nan Yang and Xiaolong Huang and Linjun Yang and Rangan Majumder and Furu Wei},
  title     = {Improving Text Embeddings with Large Language Models},
  booktitle = {Proceedings of the 62nd Annual Meeting of the Association for Computational Linguistics (Volume 1: Long Papers)},
  pages     = {11897--11916},
  year      = {2024},
  doi       = {10.18653/v1/2024.acl-long.642},
  url       = {https://doi.org/10.18653/v1/2024.acl-long.642}
}

@article{azizi2025graph,
  author    = {Ilias Azizi and Karima Echihabi and Themis Palpanas},
  title     = {Graph-Based Vector Search: An Experimental Evaluation of the State-of-the-Art},
  journal   = {Proceedings of the ACM on Management of Data},
  volume    = {3},
  number    = {1},
  note      = {Art. no.~43, 31~pages},
  year      = {2025},
  doi       = {10.1145/3709693},
  url       = {https://doi.org/10.1145/3709693}
}

@inproceedings{ootomo2024cagra,
  author    = {Hiroyuki Ootomo and Akira Naruse and Corey Nolet and Ray Wang and Tamas Feher and Yong Wang},
  title     = {{CAGRA}: Highly Parallel Graph Construction and Approximate Nearest Neighbor Search for {GPUs}},
  booktitle = {2024 IEEE 40th International Conference on Data Engineering (ICDE)},
  pages     = {4236--4247},
  year      = {2024},
  doi       = {10.1109/ICDE60146.2024.00323},
  url       = {https://doi.org/10.1109/ICDE60146.2024.00323}
}

@article{voruganti2025mirage,
  author    = {Sairaj Voruganti and M. Tamer {\"O}zsu},
  title     = {{MIRAGE-ANNS}: Mixed Approach Graph-based Indexing for Approximate Nearest Neighbor Search},
  journal   = {Proceedings of the ACM on Management of Data},
  volume    = {3},
  number    = {3},
  note      = {Art. no.~188, 27~pages},
  year      = {2025},
  doi       = {10.1145/3725325},
  url       = {https://doi.org/10.1145/3725325}
}

@article{gti2024,
  author    = {Ruiyao Ma and Yifan Zhu and Baihua Zheng and Lu Chen and Congcong Ge and Yunjun Gao},
  title     = {{GTI}: Graph-based Tree Index with Logarithm Updates for Nearest Neighbor Search in High-Dimensional Spaces},
  journal   = {Proceedings of the VLDB Endowment},
  volume    = {18},
  number    = {4},
  pages     = {986--999},
  year      = {2024},
  doi       = {10.14778/3717755.3717760},
  url       = {https://doi.org/10.14778/3717755.3717760}
}

@article{wolverine2025,
  author    = {Dawei Liu and Bolong Zheng and Ziyang Yue and Fuhao Ruan and Xiaofang Zhou and Christian S. Jensen},
  title     = {{Wolverine}: Highly Efficient Monotonic Search Path Repair for Graph-based {ANN} Index Updates},
  journal   = {Proceedings of the VLDB Endowment},
  volume    = {18},
  number    = {7},
  pages     = {2268--2280},
  year      = {2025},
  doi       = {10.14778/3734839.3734860},
  url       = {https://doi.org/10.14778/3734839.3734860}
}

@inproceedings{prohnsw2026,
  author    = {Huijun Jin and Jieun Lee and Shengmin Piao and Sangmin Seo and Sanghyun Park},
  title     = {{PRO-HNSW}: Proactive Repair and Optimization for High-Performance Dynamic {HNSW} Indexes},
  booktitle = {42nd IEEE International Conference on Data Engineering (ICDE 2026)},
  year      = {2026},
  url       = {https://icde2026.github.io/accepted-papers.html}
}

@article{greator2025,
  author    = {Song Yu and Shengyuan Lin and Shufeng Gong and Yongqing Xie and Ruicheng Liu and Yijie Zhou and Ji Sun and Yanfeng Zhang and Guoliang Li and Ge Yu},
  title     = {A Topology-Aware Localized Update Strategy for Graph-Based {ANN} Index},
  journal   = {Proceedings of the VLDB Endowment},
  volume    = {19},
  number    = {3},
  pages     = {495--508},
  year      = {2025},
  doi       = {10.14778/3778092.3778108},
  url       = {https://doi.org/10.14778/3778092.3778108}
}

@article{conda2026,
  author    = {Darae Lee and Min-Soo Kim},
  title     = {{CONDA}: A Connectivity-Aware Dynamic Index for Approximate Nearest Neighbor Search over Evolving Data},
  journal   = {Proceedings of the VLDB Endowment},
  volume    = {19},
  number    = {11},
  pages     = {3357--3370},
  year      = {2026},
  doi       = {10.14778/3836663.3836694},
  url       = {https://doi.org/10.14778/3836663.3836694}
}

@inproceedings{odinann2026,
  author    = {Hao Guo and Youyou Lu},
  title     = {{OdinANN}: Direct Insert for Consistently Stable Performance in Billion-Scale Graph-Based Vector Search},
  booktitle = {24th USENIX Conference on File and Storage Technologies (FAST 26)},
  pages     = {133--147},
  year      = {2026},
  url       = {https://www.usenix.org/conference/fast26/presentation/guo}
}

@inproceedings{mlembed2026,
  author    = {Ziyin Zhang and Zihan Liao and Hang Yu and Peng Di and Rui Wang},
  title     = {{ML-Embed}: Inclusive and Efficient Embeddings for a Multilingual World},
  booktitle = {Proceedings of the 43rd International Conference on Machine Learning},
  series    = {Proceedings of Machine Learning Research},
  volume    = {306},
  pages     = {157026--157063},
  year      = {2026},
  url       = {https://proceedings.mlr.press/v306/zhang26dh.html}
}

@inproceedings{matryoshkaadaptor2024,
  author    = {Jinsung Yoon and Rajarishi Sinha and Sercan O. Arik and Tomas Pfister},
  title     = {{Matryoshka-Adaptor}: Unsupervised and Supervised Tuning for Smaller Embedding Dimensions},
  booktitle = {Proceedings of the 2024 Conference on Empirical Methods in Natural Language Processing},
  pages     = {10318--10336},
  year      = {2024},
  doi       = {10.18653/v1/2024.emnlp-main.576},
  url       = {https://doi.org/10.18653/v1/2024.emnlp-main.576}
}

@inproceedings{continualcompatible2024,
  author    = {Zhenyu Cui and Jiahuan Zhou and Xun Wang and Manyu Zhu and Yuxin Peng},
  title     = {Learning Continual Compatible Representation for Re-indexing Free Lifelong Person Re-identification},
  booktitle = {Proceedings of the IEEE/CVF Conference on Computer Vision and Pattern Recognition (CVPR)},
  pages     = {16614--16623},
  year      = {2024},
  url       = {https://openaccess.thecvf.com/content/CVPR2024/html/Cui_Learning_Continual_Compatible_Representation_for_Re-indexing_Free_Lifelong_Person_Re-identification_CVPR_2024_paper.html}
}

@inproceedings{crossmodalcompatible2025,
  author    = {Young Kyun Jang and Ser-Nam Lim},
  title     = {Towards Cross-modal Backward-compatible Representation Learning for Vision-Language Models},
  booktitle = {Proceedings of the IEEE/CVF International Conference on Computer Vision (ICCV)},
  pages     = {1783--1792},
  year      = {2025},
  url       = {https://openaccess.thecvf.com/content/ICCV2025/html/Jang_Towards_Cross-modal_Backward-compatible_Representation_Learning_for_Vision-Language_Models_ICCV_2025_paper.html}
}

@inproceedings{driftadapter2025,
  author    = {Harshil Vejendla},
  title     = {{Drift-Adapter}: A Practical Approach to Near Zero-Downtime Embedding Model Upgrades in Vector Databases},
  booktitle = {Proceedings of the 2025 Conference on Empirical Methods in Natural Language Processing},
  pages     = {15938--15949},
  year      = {2025},
  doi       = {10.18653/v1/2025.emnlp-main.805},
  url       = {https://doi.org/10.18653/v1/2025.emnlp-main.805}
}

@inproceedings{embeddingconverter2025,
  author    = {Jinsung Yoon and Sercan O. Arik},
  title     = {{Embedding-Converter}: A Unified Framework for Cross-Model Embedding Transformation},
  booktitle = {Proceedings of the 63rd Annual Meeting of the Association for Computational Linguistics (Volume 1: Long Papers)},
  pages     = {25464--25482},
  year      = {2025},
  doi       = {10.18653/v1/2025.acl-long.1237},
  url       = {https://doi.org/10.18653/v1/2025.acl-long.1237}
}

@article{wang2025flash,
  author    = {Mengzhao Wang and Haotian Wu and Xiangyu Ke and Yunjun Gao and Yifan Zhu and Wenchao Zhou},
  title     = {Accelerating Graph Indexing for {ANNS} on Modern {CPUs}},
  journal   = {Proceedings of the ACM on Management of Data},
  volume    = {3},
  number    = {3},
  note      = {Art. no.~123, 29~pages},
  year      = {2025},
  doi       = {10.1145/3725260},
  url       = {https://doi.org/10.1145/3725260}
}

@article{yang2025revisiting,
  author    = {Shuo Yang and Jiadong Xie and Yingfan Liu and Jeffrey Xu Yu and Xiyue Gao and Qianru Wang and Yanguo Peng and Jiangtao Cui},
  title     = {Revisiting the Index Construction of Proximity Graph-Based Approximate Nearest Neighbor Search},
  journal   = {Proceedings of the VLDB Endowment},
  volume    = {18},
  number    = {6},
  pages     = {1825--1838},
  year      = {2025},
  doi       = {10.14778/3725688.3725709},
  url       = {https://doi.org/10.14778/3725688.3725709}
}

@inproceedings{platonic2024,
  author    = {Minyoung Huh and Brian Cheung and Tongzhou Wang and Phillip Isola},
  title     = {Position: The {Platonic} Representation Hypothesis},
  booktitle = {Proceedings of the 41st International Conference on Machine Learning},
  series    = {Proceedings of Machine Learning Research},
  volume    = {235},
  pages     = {20617--20642},
  year      = {2024},
  url       = {https://proceedings.mlr.press/v235/huh24a.html}
}
\end{document}